\documentclass[10pt]{article}

\usepackage{hyperref}
\newcommand{\footremember}[2]{%
    \footnote{#2}
    \newcounter{#1}
    \setcounter{#1}{\value{footnote}}%
}
\newcommand{\footrecall}[1]{%
    \footnotemark[\value{#1}]%
}

\date{}

\usepackage{graphicx}

\title{Parallel photonic reservoir computing using frequency multiplexing of neurons}

\author{Akram Akrout\footremember{LIQ} { Laboratoire d'Information Quantique, CP 224, Universit\'e libre de Bruxelles (U.L.B.), Boulevard du Triomphe, 1050 Brussels, Belgium}
\and
Arno Bouwens\footremember{OPERA}{OPERA-Photonique, CP 194/5, Universit\'e libre de Bruxelles (U.L.B.), 
Avenue Adolphe Buyl 87, 1050 Brussels, Belgium}
\and
 Fran\c cois Duport\footrecall{OPERA} 
 \and
Quentin Vinckier \footrecall{OPERA} 
\and
Marc Haelterman\footrecall{OPERA} 
\and 
Serge Massar\footrecall{LIQ}
} 

\begin{document}

\maketitle

\begin{abstract}
Today's unrelenting increase in demand for information processing creates the need for novel computing concepts. Reservoir computing is such a concept that lends itself particularly well to photonic hardware implementations. Over recent years, these hardware implementations have gained maturity and now achieve state-of-the-art performance on several benchmark tasks. However, implementations so far are essentially all based on sequential data processing, leaving the inherent parallelism of photonics unexploited. Parallel implementations process all neurons simultaneously, and therefore have the potential of reducing computation time by a factor equal to the number of neurons, compared to sequential architectures. Here, we report a parallel reservoir computer that uses frequency domain multiplexing of neuron states. We illustrate its performance on standard benchmark tasks such as nonlinear channel equalization, the reproduction of a nonlinear 10th-order system, and speech recognition, obtaining error rates similar to previous optical experiments. The present experiment is thus an important step towards high speed, low footprint, all optical photonic information processing.
\end{abstract}

\maketitle



\section{Introduction}

\begin{figure}[htbp]
\centering\includegraphics[width=9cm]{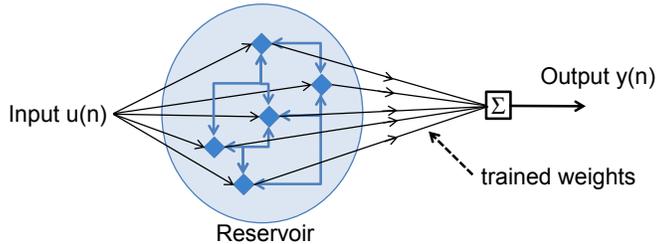}
\caption{Schematic of a generic reservoir computer. The reservoir layer is a nonlinear recurrent dynamical system, driven by the input u(n), with n denoting discrete time. The output y(n), produced in the output layer, is a linear combination of the internal variables. Only the linear output layer is adjusted during the training phase.}
\label{Fig1A}
\end{figure}

Reservoir computing (RC) is a bio-inspired computational paradigm\cite{1,2,3}, that has been shown to perform remarkably well in the field of time-dependent signal processing. It has been applied to tasks such as speech recognition, nonlinear channel equalization, detection of epileptic seizures, time series prediction, handwriting recognition, etc, see the review articles\cite{4,5}. A reservoir computer is made of a large number of interconnected nonlinear nodes, also called neurons or internal variables. This network, the reservoir, contains feedback connections and therefore behaves as a recurrent dynamical system. It is driven by a time dependent input signal, as shown in Figure \ref{Fig1A}. The output layer collects the reservoir's time-dependent response to the input signals via a weighted sum of the individual node states. The reservoir computer is trained to perform a signal processing task by optimizing the weights of the output layer. In addition, tuning some global system parameters can further optimize the reservoir computer's performance. Unlike a traditional neural network, however, the interconnection weights between nodes in a reservoir computer are not tuned individually, which greatly simplifies the training procedure.

During the last decade, several physical implementations of RC have been reported\cite{6,7,8,9,10,11,12,13,14,15}. Of particular interest are the photonics-based reservoir computers, thanks to the remarkable speed and multiplexing capabilities of optical components. These were first proposed in\cite{16,17}, and a number of experimental demonstrations have since been reported\cite{8,9,10,11,12,13,14,15}, see also the numerical simulations of\cite{18,19,20,21}. Many of these experiments are based on a specific architecture consisting of a single nonlinear node with delayed feedback, first introduced using an electronic system in\cite{7}, see also the related theoretical proposal\cite{22,23}. In these systems, the delayed feedback is realized by a fiber loop and neurons propagate sequentially through the loop, amounting to time domain multiplexing of the neuron states. With this approach it is possible to build photonic reservoir computers that can solve, with error rates comparable to digital implementations, real world tasks such as speech recognition, time series prediction, channel equalization. 

Recently, reservoir computers based on a linear optical network have been studied experimentally\cite{12,13}, see also the related theoretical proposal\cite{24,25}. In these systems, the nonlinearity required for reservoir computing is realised by the quadratic dependence of the current generated in the read-out photodiode on the amplitude of the optical electric field. In this way, the nonlinearity is created in the read-out layer instead of the reservoir layer. These systems employ coherent light, i.e. encoding information in both phase and amplitude of the injected field, which can significantly improve performance with respect to systems using only intensity modulation. This was first shown through numerical analysis in\cite{17}, and then experimentally in\cite{13} which combined coherent coding and a delay loop to obtain record low error rates on a number of tasks.
Thus, today, the experimental reservoir computers based on time domain multiplexing can solve the most complex tasks, and achieve the lowest error rates on benchmark tasks. However, these systems suffer from an inherent trade-off between the number of neurons in the system and the required processing time. Indeed, adding more neurons to a time domain multiplexed system requires either increasing the length of the feedback loop or increasing the modulation frequency of the input signal. Because the input signal's modulation frequency is ultimately limited by the input and read-out electronics, the only remaining solution is to lengthen the feedback loop, and thus increase the total processing duration.

However light has multiple degrees of freedom: polarisation, frequency, space. By multiplexing the neurons in one of these domains, one could circumvent the above tradeoffs, and process multiple neurons simultaneously. One can take inspiration from optical telecommunications, where multiplexing in multiple degrees of freedom is widely used to increase the bandwidth of optical fibers. 
This leads to an important distinction between sequential (or time multiplexed) and parallel reservoir computer architectures. More precisely, operating a reservoir computer requires many different operations such as: adding the current input to the internal states, linearly combining the internal states, carrying out a nonlinear operation on the internal states, multiplying the internal states by the output weights, summing to obtain the output (see Figure \ref{Fig1A}). Denote the time taken for the slowest of these operations by $\theta$. Then a for a sequential reservoir with $N$  internal variables, the time after which the input can be updated will be $N\theta$. For instance in\cite{14} (the fastest sequential reservoir computer reported so far), $\theta=200$ps , $N=388$, and the input update rate was $N\theta=77,6$ns. On the contrary, in a parallel architecture the input update time will be equal to $\theta$ independently of $N$, leading to a considerable gain in overall operating speed. (Of course as one increases the number of internal states $N$, the footprint of the reservoir will start to increase, and at some point the input update time will need to be increased. But the scaling will necessarily be much better than for a sequential reservoir). For this reason, the distinction between parallel and sequential reservoir computers is a very important one, and we expect much of the future effort in this area to focus on parallel architectures.

In \cite{12}, Vandoorne et al. demonstrated a photonic reservoir computer in which neurons were implemented as physically separate delay lines on a silicon chip, i.e. they were spatially multiplexed. This can be considered as the first demonstration of a parallel optical reservoir computer. Although their approach is very promising, some drawbacks can be identified: the coupling between neurons and the number of neurons is fixed by the chip's topology, and the chip's size increases with the number of neurons. Furthermore the tasks on which it was possible to test this system were significantly simpler than those that have been solved by delay reservoirs. Demonstrating the full potential of parallel optical reservoir computing thus remains one of the major open challenges in the field.  

\begin{figure}[htbp]
\centering\includegraphics[width=12cm]{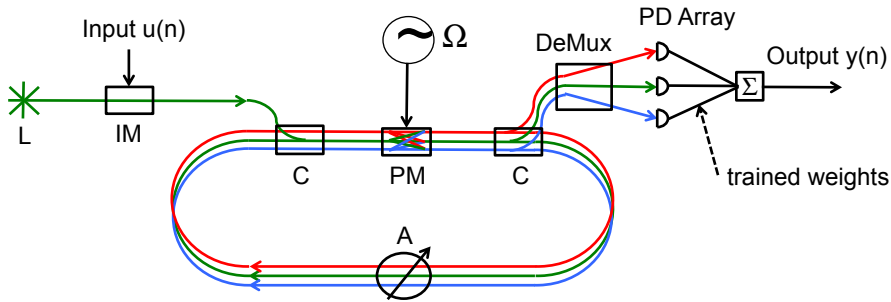}
\caption{Schematic of a photonic reservoir computer based on frequency parallelism. The input u(n) is encoded in the amplitude of the electromagnetic field of a monochromatic Laser (L) using an Intensity Modulator (IM). The reservoir consists of a fiber loop in which the different frequencies propagate in parallel and coherently mix. The input is injected into the reservoir using a coupler (C). The different frequencies are mixed using a phase modulator (PM) driven by a periodic Radio Frequency signal at frequency $\Omega/2\pi$. A tunable Attenuator (A) allows adjusting the feedback within the reservoir. The different frequencies are coupled out of the reservoir using a second coupler (C), and separated using a Frequency Demultiplexer (DeMux). The amplitudes converted to electrical currents using an array of photodiodes (PD array). The output consists of a linear combination of these electrical currents. Only the output weights are adjusted during the training phase. The nonlinearity in this architecture is given by the readout photodiodes whose response is quadratic (current proportional to the norm square of the electric field amplitude). Note that in the experiment reported here, a single readout photodiode was used, rather than an array, and the different frequencies (neurons) were read out sequentially by repeating the experiment.}
\label{Fig1B}
\end{figure}

Here, we present a proof of principle parallel photonic reservoir computer whose performance on benchmark tasks is comparable to reservoirs based on delay loops. In our experiment the reservoir states are coded in the amplitude and phase of the frequency sidebands of a narrow band laser, propagating in a single-mode fiber optical loop. The quadratic non-linearity is provided by the readout photodiode. The sidebands are created and manipulated using an electro-optic phase modulator driven by a Radio Frequency (RF) signal, following ideas and techniques introduced in quantum optics\cite{26,27}. A schematic of the operating principle is given in Figure \ref{Fig1B}. We evaluate performance on three benchmark tasks often used in the RC community: reproduction of a nonlinear auto regressive moving average model (NARMA10), isolated spoken digit recognition, and nonlinear channel equalization, obtaining results comparable to the state of the art. The present results thus constitute an essential step on the path towards development of high performance high-speed parallel optical reservoir computers.  Due to technical limitations, in the present experiment only $13$ neurons are processed in parallel, and the refresh rate of the input is a comparatively slow $12.5$ MHz. In the conclusion we discuss how these limitations could be overcome in future work.

\section{Operating Principle}
The input of the reservoir is a time series $u(n)$, rescaled to belong to the interval $[-1,+1]$, with $n\in N$ denoting discrete time. We transform this into a continuous time input using a sample and hold procedure $u(t)=u(n)$ when $t\in[ n\tau,(n+1)\tau[$, with $t\in \Re$ denoting continuous time, and $\tau$ the hold time, equal to the roundtrip time of the cavity (see below). The light source emits a continuous wave (CW) narrow band laser at frequency $\omega_0$ and amplitude $E$ propagating in a single-mode fiber. After the input is encoded, the light amplitude at the input of the reservoir is $E u(t)e^{-j\omega_0t}$, where $j^2=-1$ . This input signal is injected into a single-mode fiber cavity which represents the reservoir. Inside the cavity, a Phase Modulator (PM) driven by a RF signal at frequency $\Omega\ll \omega_0$ creates frequency sidebands. Specifically the phase modulator acts on a light amplitude at frequency $\omega$ as
\begin{equation}
Ee^{-j\omega t}\to Ee^{-j\omega t}e^{+j m \sin{\Omega t}}=\sum_k E J_k(m) (-1)^k e^{-j(\omega +k\Omega)t}
\label{Eq1}
\end{equation}
where $m$ is the amplitude of the modulation (experimentally we are able to reach up to $m=2.0$), and $J_k (m)$ is the kth-order Bessel function of the first kind, and we have used the Jacobi-Anger expansion. 

The field amplitude inside the cavity can  be written as 
$\sum_k x_k(t) e^{-j(\omega +k\Omega)t}$ 
where $x_k (t)$ are slowly varying amplitudes, approximately constant over the time intervals of duration $\tau$. We can thus introduce the discrete time amplitudes $x_k (n)=x_k (t)$, when $t\in [n\tau,(n+1)\tau[$, and from eq. (\ref{Eq1}) we can write their dynamics as
\begin{equation}
x_k (n+1)=\alpha e^{j\varphi(k)} \sum_l J_l (m) (-1)^l x_{k-l} (n) +\beta u(n) \delta_(k,0)   
\label{Eq2}
\end{equation}
where $\alpha$ is the feedback strength, $\beta$ is the input strength, and the Kronecker delta function $\delta_(k,0)$ encodes the fact that the input is encoded only on the central frequency $k = 0$. The phase accumulated by the sideband at frequency $\omega_0+k\Omega$ over one roundtrip of the cavity detuning is denoted $\varphi(k)$. Considering the relatively short cavity length one can neglect group-velocity dispersion in the fiber and this function of $k$ can therefore be approximated by its first-order Taylor expansion $\varphi(k)=\varphi_0+k\varphi_1$. 
Experimentally $\varphi_0$ can be readily tuned while, for a given fiber, $\varphi_1$ is fixed by the central frequency $\omega_0$ and the modulation frequency $\Omega$. 
Part of the light amplitude is sent to the readout layer in which a bandpass filter selects the frequency bands centered on $\omega_0+k\Omega$ and sends the corresponding fields to the photodiodes that produce the electric currents proportional to $|x_k (n)|^2$. The reservoir output $y(n)$ is thus obtained by performing a linear combination of these electric currents : 
\begin{equation}
y(n)=\sum_{i=0}^{N-1} w_i |x_i (n)|^2
\label{Eq3}
\end{equation}
where $w_i$ are the readout weights. Note that in the framework of the present proof of principle experiment we use only one photodiode and the reservoir output $y(n)$ is built on the basis of the recorded signals  $|x_k (n)|^2$ obtained through successive identical runs. 
During the training phase, the reservoir computer is fed with an input signal for which the target output $\tilde y(n)$ is already known. Once the values of $|x_k (n)|^2$ are recorded, we calculate the read-out weights that minimize the Normalized Mean Square Error (NMSE) between the reservoir output $y(n)$ and the target output $\tilde y(n)$:
\begin{equation}
NMSE=  \frac{\langle (\tilde y(n)-y(n))^2 \rangle_n}{\langle (\tilde y(n)-\langle \tilde y(n)\rangle_n )^2 \rangle_n}
\label{Eq4}
\end{equation}
where$\langle \rangle_n$ denotes the average over the discrete time steps. Ridge regularisation may be used at this step to avoid overfitting to the training data. The training step is repeated to find optimal values for the parameters $\alpha$, $\varphi_0$, $m$. (Note that since the reservoir is linear and the output nonlinearity is quadratic, the input amplitude $\beta$ can be chosen arbitrarily).
In the test phase, the readout weights $w_i$ are kept fixed, a new input signal is fed into the reservoir, and the output $y(n)$ is obtained using Eq. (\ref{Eq3}). The output is compared to the target output $\tilde y(n)$ to evaluate the performance of the reservoir computer.

\section{Experimental implementation}

\begin{figure}[htbp]
\centering\includegraphics[width=9cm]{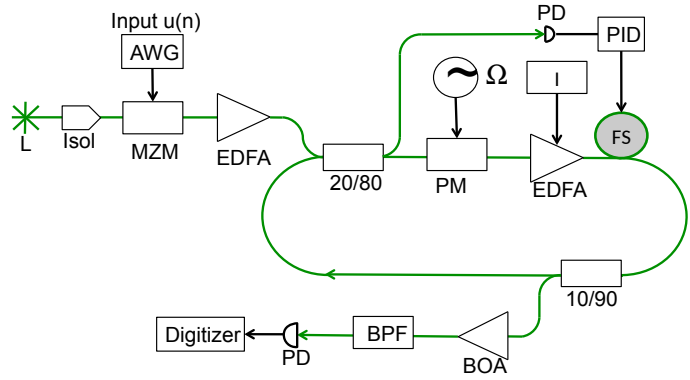}
\caption{Schematic view of the experimental reservoir computer. Green lines are polarisation maintaining optical fiber; black lines are electrical connections. L: narrowband laser, Isol: Isolator, EDFA: Erbium-doped fiber amplifier, MZM: Mach-Zender intensity modulator, AWG: Arbitrary Waveform Generator, PM: phase modulator, I: tunable current source, FS: fiber stretcher, PID: proportional-integral-derivative controller, BOA: Booster Optical Amplifier, PD: photodiode, BPF: programmable Band Pass Filter (Finisar Waveshaper). Note that in the experiment, we choose which frequency bin (i.e. neuron to measure) by tuning the BPF, record its time trace, and then repeat the experiment with another frequency bin selected.  }
\label{Fig2}
\end{figure}

\begin{figure}[htbp]
\centering\includegraphics[width=8cm]{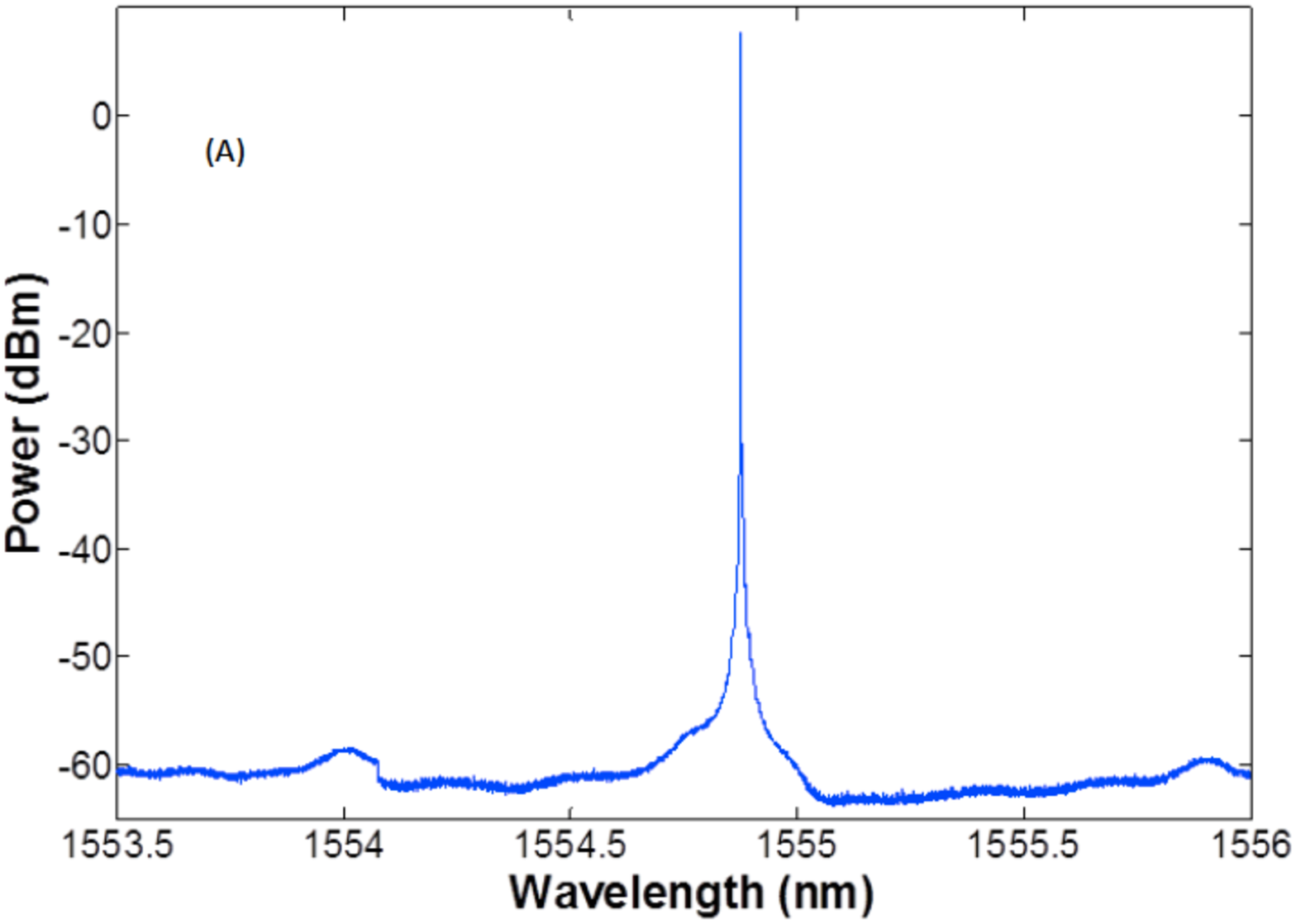}
\centering\includegraphics[width=8cm]{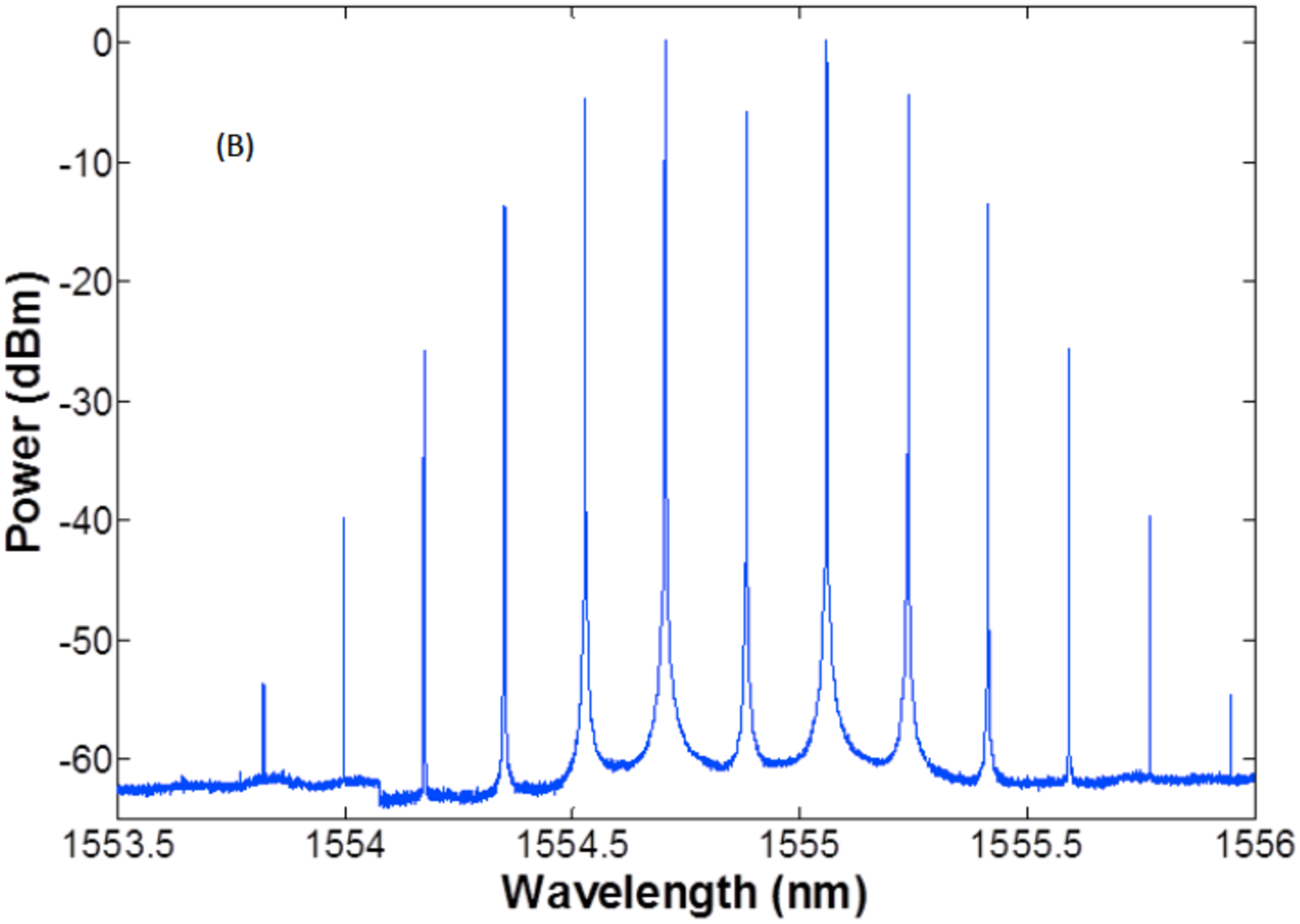}
\centering\includegraphics[width=8cm]{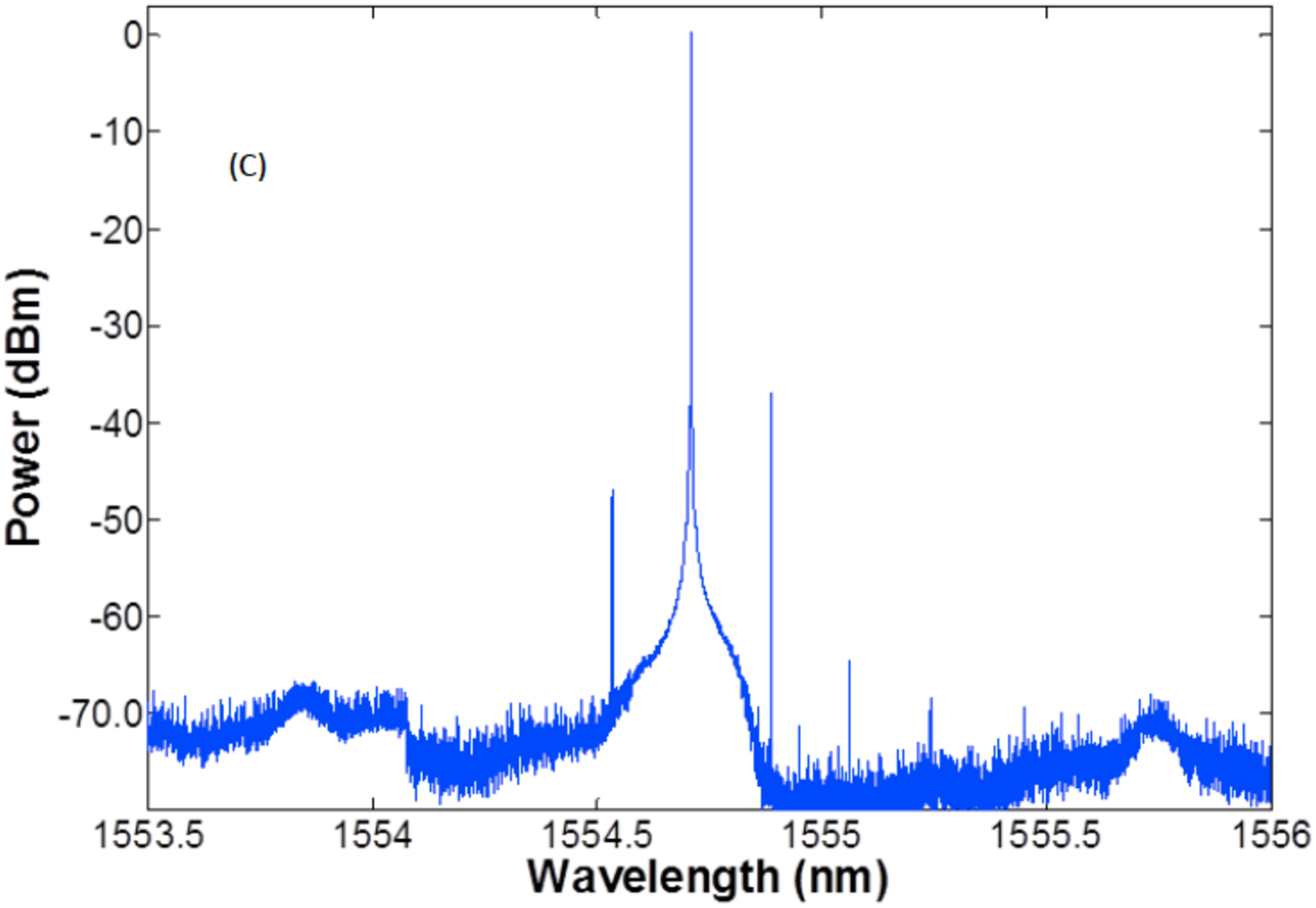}
\caption{Spectra. (A) Spectrum of the laser. (B) Spectrum inside the cavity. In this spectrum 11 frequency sidebands are visible. In the experiment, 13 sidebands were used. (C) Spectrum after a specific frequency bin (i.e. neuron) is selected by the BPF. Spectra (A) and (B) are obtained without the BOA whereas spectrum (C) constitutes an example of measured signal obtained after amplification by the BOA. }
\label{Fig2BCD}
\end{figure}

Our optical experiment, depicted in Figure \ref{Fig2}, uses polarisation maintaining fiber and operates in the telecom C-band. The optical signal is generated by a continuous wave (CW) laser at wavelength $\lambda_0=2\pi c/\omega_0=1555$nm with a coherence time much greater than the inverse linewidth of the cavity ($c$ is the speed of light in vacuum). The laser spectrum is depicted in Figure \ref{Fig2BCD} (A). The input signal is then encoded with a Mach-Zenhder (MZ) Lithium Niobate intensity modulator driven by an arbitrary waveform generator (AWG). Since the MZ modulator has a sinusoidal transfer function, we tried to precompensate the voltage V generated by the AWG in order to get an input signal proportional to $u(t)$. We also omitted this pre-compensation whereupon the optical signal at the output of the MZ is proportional to  $\sin[V(t)\pi/ (2V_{\pi,RF})]$ with $V(t)=\beta u(t)\in [-\beta V_{\pi,RF}, \beta V_{\pi,RF} ]$ ($V_{\pi,RF}$  is the MZ characteristic voltage and $\beta\in[0,1]$. In our case, both coding scenarios allow to obtain the same performances. We have also experimentally verified that for some tasks, it is necessary to bias the input. Therefore, a direct current (DC) voltage $V_(0 )\in[-V_{\pi,DC},V_{\pi,DC} ]$ is applied to the DC electrode of the MZ. Then, the input signal of the cavity is proportional to $\sin[(\pi/2) (V(t)/V_{\pi,RF} +V_0/V_{\pi,DC} )]$.The resulting signal is injected into the cavity using an $80/20$ fiber coupler. The power injected into the cavity after the $80/20$ fiber coupler can be adjusted in the range $8$ to $63$ mW. The roundtrip time of the cavity is $\tau=81.17$ ns, corresponding to a length of $\approx 17$ m. 

Inside the cavity, a phase modulator driven by a RF signal at frequency $\Omega/2\pi=21$GHz creates frequency sidebands.  The voltage characteristic of the PM is $V_\pi=2.7$ V up to about $\Omega/2\pi=20$ GHz when it starts to increase. The chosen frequency $\Omega/2\pi=21$GHz 
is a compromise: the higher $\Omega$, the easier it is to separate the sidebands afterwards, but above a certain value of $\Omega$ the efficiency of the phase modulator decreases. The chosen frequency 
is the maximum that allows the creation of $13$ frequency sidebands, obtained for a modulation amplitude $m=2.0$. The  spectrum inside the cavity, when the phase modulator creates sidebands, is depicted in Figure \ref{Fig2BCD} (B).

In order to reach good performance, the feedback strength $\alpha$ should be tuned to be close to $1$. An Erbium Doped Fiber Amplifier (EDFA) is inserted in the cavity to compensate the $\approx 4$dB phase modulator optical losses.  Adjusting the pump current of the EDFA provides a simple way to adjust the feedback strength $\alpha$.

The roundtrip phase $\varphi(k)=\varphi_0+k\varphi_1$ is an important parameter in our experiment. In order to keep $\varphi_0$ and $\varphi_1$ fixed, the $\approx 17$m delay loop needs to be stabilized. To this end, a proportional-integral-derivative (PID) controller drives a piezoelectric fiber stretcher inserted in the cavity. The optical intensity at the output of the interferometer serves as the control signal for the PID. This stabilization system has been used and detailed previously by our group\cite{13}. The latter implementation employs two counter-propagating modes in the fiber cavity: one mode carries the information processed by the reservoir, while the other mode serves as the PID's control signal.  However, our present setup supports only one propagation direction, as the intra-cavity EDFA amplifier includes an isolator. Therefore, a single propagation direction is used to carry both the information to be processed and the control signal, but in a sequential manner. Each time data is being processed by the reservoir, the PID controller is disabled temporarily and the cavity length is kept constant. Indeed, we experimentally verified that, during the time needed to process one set data ($\simeq 0.8$ms), the roundtrip phase $\varphi(k)$ (or cavity detuning) remains constant. The above procedure allows us to keep $\varphi_0$ fixed, and to scan its value for optimal performance. 

Part of the intra-cavity signal is sent to the output using a $90/10$ coupler. A band-pass filter selects one of the frequency sidebands. The corresponding amplitude is sent to a photodiode after being amplified using a Booster Optical Amplifier (BOA) to increase the SNR (minimization of photodiode noise contribution). The  output spectrum, after filtering and amplification by the BOA, is depicted in Figure \ref{Fig2BCD} (C).
The resulting electrical signal is recorded by an oscilloscope and processed on a computer. In order to further increase the SNR, the neuron values are recorded $10$ times, and the average of these $10$ acquisitions used in the post-processing. (The reason why this averaging is necessary is that some of the frequency sidebands are very weak, see Figure \ref{Fig2BCD} (B). If a frequency comb was used as source instead of a monochromatic laser, the intensities of the neurons would be more uniform, the SNR consequently higher, and averaging not necessary). 

The output sequence $y(n)$ is then computed offline using the external computer Eq. (\ref{Eq3}). To this end, the entire intensity trace for each frequency neuron is rescaled to lie between $-1$ and $+1$. Then each internal state is measured by taking a temporal average of a window of length $\tau/2$ centered on the middle of each window of length $\tau$. In this way we minimize the influence of the oscilloscope and photo-detector intrinsic noise on the internal state values. The whole data acquisition is repeated for each frequency sideband, by adjusting the bandpass filter. This approach allows us to demonstrate the performance and feasibility of reservoir computing with frequency-multiplexed neurons, without requiring a dedicated photodiode per neuron.

The input sequence is divided in three subsequences. We first send a ``warm up'' sequence to eliminate any memory effects within the reservoir, followed by the training input sequence, and finally the test sequence. The output weights $w_i$ are computed from the reservoir states recorded during the training phase using a least mean square algorithm that minimizes the NMSE defined in Eq. (\ref{Eq4}). Finally the performance of the reservoir is computed by evaluating its performance on the test sequence, using the readout weights determined from the training sequence. 
Further details on the experimental setup, including detailed description of the components used, are given in Table 1.

\begin{table}[htp]
\caption{Characteristics of hardware components used in the experiment.}
\begin{center}
\begin{tabular}{|c|c|}
\hline
Equipment & Characteristics \\
\hline
Laser &	
NKT Photonics: Koheras Basik Module Fiber laser,\\
&Model: K82-152-13. SNR: 60 dB,\\
& Wavelength: 1555 nm, Linewidth :  100 Hz, Output power : [8, 200] mW.
\\
\hline
Mach-Zehnder Modulator	&
Photline electro-optic modulator. Model: MX-LN-10\\
&Bandwidth : 12 GHz, Vpi RF (@ 50kHz) : 4.5 V, \\
&Extinction ratio : 30 dB, Insertion loss : 3.5 dB.\\
\hline
Phase Modulator & EOspace Ultra-Low Vpi phase modulator:\\
& 3 dB Bandwidth :     25 GHz, Vpi (@ 1 GHz) : 2.7 Volts,\\
&Optical return loss : > 50 dB, Insertion loss : 4 dB.\\
\hline
Arbitrary waveform
generator	& NI PXI 5422. 16 Bit 200 MS/s AWG.\\
\hline
Photodiodes	& TTI-TIA-525. 120 MHz bandwidth.\\
\hline
Acquisition system	&National Instrument NIPXI-5124 Digitizer.\\
&12 Bits, up to 200 MSamples/s per channels.\\
\hline
PID regulator	& Homemade PID regulator\\ 
&programmed
using mbed NXP LPC1768 development
board:\\
&-Processor: 32-bits ARM Cortex-M3 core
running at 96 MHz,\\
&-Analog-to-digital converter (ADC):
12 bits resolution,\\
&Digital-to-analog converter (DAC):
10 bits resolution,\\
&Output refresh rate: 8.3 kHz.\\
\hline
Band pass Filter & Finisar waveshaper 4000s:\\
&Operating Frequency Range \\
& 191.250 THz 
to 196.275 THz (1527.4 nm to 1567.5 nm),\\
&Insertion Loss ~ 4.5 to 6.5 dB, Filter Bandwidth : 10 GHz - 5 THz,\\
&Center Frequency Setting Resolution : 1 GHz (8 pm).\\
\hline
Boost Optical Amplifier & Thorlabs, BOA 1004 P, semiconductor
 based amplifier:\\
&Wavelength range: 1530-1570 nm\\
&Output power : 15 dBm, Gain: 27 dB .\\
\hline
Erbium Doped Fiber amplifiers & Keopsys, CEFA-C-BO-HP-PM:\\
(inside the
cavity) &Wavelength range: 1540-1565 nm,\\
&Output power : 27 dBm, Input power range: 5-15 dBm.\\
\hline
Erbium Doped Fiber amplifiers & Keopsys, KPS-CUS-BT-C:\\
(input) & Wavelength range: 1535-1565 nm,\\
& Output power : 25 dBm, Input power range: -15-0 dBm.\\
\hline
\end{tabular}
\end{center}
\label{default}
\end{table}%

\section{Results}

We first developed numerical models for the experiment. A discrete-time model based on numerically integrating the recurrence Eq. (\ref{Eq2}) provides a first approximation of the experiment. This discrete-time numerical model was used initially to validate our approach. Then, a more accurate continuous-time numerical model was developed, taking into account the transfer functions of all passive and active components that were used. The goal of this model was to verify the feasibility of the experiment, and to investigate the performance of the system as a function of its global parameters. Indeed, many such global parameters can be scanned to optimize our reservoir's performance for a given task ($\alpha$, $\varphi_0$, $m$, $\Omega$, and $V_0$) and the ridge parameter). However, running the physical experiment for all combinations of parameters is too time-consuming, hence the simulations serve to provide a subset of parameter ranges to be tested in the physical experiment. 

To test our device, we evaluated its memory capacities, as well as its performance on three benchmark tasks often considered in the RC community. 
In order to validate the performance of the experiment, we compared it with the continuous-time model. The present experiment has only $13$ measurable frequency bands. In order to see whether any limitation of performance is linked to the principle of the experiment, or is due to the limited number of neurons, we artificially increased the number of neurons in simulation by increasing the modulation amplitude $m$ until we had $51$ neurons, which makes our system comparable to existing ones. 

\subsection{Memory capacities of the reservoir}

The reservoir memory capacities is a basic task that simply evaluates the ability of the reservoir to process linear and nonlinear functions of its past inputs $u(n-k)$. The inputs, used for this task, are independent and identically distributed random variables in the interval $[-1,+1]$. The capacity of the reservoir (CR) to compute a function $\tilde y_k (n)$ of the past inputs is defined by $CR[y_k (n)]=[1-NMSE(y_k (n))]$ \cite{28}. So that a perfect processing of $y(n)$ leads to a capacity of $1$, while a completely uncorrelated output gives $0$. In general, this capacity is measured for a set of inputs functions $y_k$ and the capacity is the summation over all function capacities: $C=\sum_k CR(y_k)$. We will consider three reservoir memory capacities; the linear, quadratic and cross memory capacities.

The reservoir linear memory capacity (LMC), introduced in\cite{1}, consist in reproducing the input $u(n-k)$ shifted $k$ time steps in the past. By summing all capacities over all delays $k$, we obtain the reservoir linear memory capacity. The linear capacity quantifies how much the reservoir remembers previous inputs.

The quadratic memory capacity (QMC) evaluates how the reservoir process the second order Legendre polynomial of the input $u$, $k$ time steps in the past: $y_k (n)=3u^2 (n-k)-1$\cite{28}. The QMC is also obtained upon summing on all values of the delay $k$. By considering the product of two inputs at two different past time steps $k$ and $k'$, $y_{kk'} (n)=u(n-k)u(n-k')$ we obtain the cross memory capacity XMC. Then by summing over all possible product couples with $k< k'$, we obtain the total cross memory capacity. 

Finally, the total memory capacity is obtained by summing the LMC, the QMC and the XMC. The total memory capacity cannot exceed the total number of internal variables in the reservoir \cite{28}.
In Table 2, we report the best experimental memory we obtained with our reservoir made of $N = 13$ internal variables, as well as the values obtained in simulation for $N=13$ and $N=51$. We compare our results with the memory capacities of our previous optoelectronic\cite{8}, SOA based all-optical\cite{11}, and the coherently driven passive reservoir computers\cite{13}. Our results were obtained for a pump current of the EDFA of $850$mA, which corresponds to a feedback gain $\alpha$ of approximately $0.81$ (this value takes into consideration all the losses in the delay loop). Results in both simulation and Experiment were obtained by scanning $\varphi_0$ in the range $[0,\pi/2]$ rad. The optical power injected into the cavity is approximately $43$mW. $10$ datasets of $3200$ inputs samples were used to evaluate the different memory capacities. 

\begin{table}[htp]
\caption{Memory capacity evaluation}
\begin{center}
\begin{tabular}{|c|c|c|c|c|}
\hline
 & Linear memory & Quadratic memory & Cross-memory & Total memory\\
 \hline
Experiment (setup under test) & 10.31 & 5.12 & 7.21& 12.32 \\
 \hline
Simulations (13 neurons) & 10.59 & 5.26 & 7.96 & 12.89  \\
 \hline
Simulations (51 neurons) & 18.81 & 8.48 & 13.74 & 30.52 \\
 \hline
Opto-electronic setup\cite{8} & 31.9 & 4.0 & 27.3  & 48.6 \\
 \hline
SOA based setup\cite{29}  & 20.84 & 4.16 & 4.71 & 25.2 \\
 \hline
Passive based cavity\cite{13}& 21.14 & 12.07 & 30.2 & 48.37 \\
 \hline
\end{tabular}
\end{center}
\end{table}%

\subsection{NARMA10}
The Nonlinear Auto Regressive Moving Average Model (NARMA) is a generalization of the Auto Regressive models and Moving Average model. In particular, NARMA10 is one of the most used benchmarks for validation of reservoir computing implementation. The aim of this task is to reproduce the nonlinear behavior of a tenth-order system using inputs $u(n)$ randomly drawn from a uniform distribution over the interval $[0, 0.5]$. That is, for a given white noise $u(n)$, the reservoir output $y(n)$  should be as close as possible to the target $\tilde y(n)$ of the NARMA10 model for the same white noise. The equation defining the target system is given by:
\begin{equation}
\tilde y(n+1)=0.3\tilde y(n)+0.05\tilde y(n)\left ( \sum_{i=0}^9 \tilde y(n-i) \right )+1.5u(n-9)u(n)+0.1           \label{Eq5}
\end{equation}

The reservoir was trained over $1000$ values of $u(n)$. Then, it was tested over $2000$ new values. The standard deviation was evaluated by repeating this procedure $10$ times. The performances obtained were measured using the NMSE. We obtained a NMSE of $0.175\pm0.03$ and $0.181\pm0.02$ for the simulation and experiment respectively. These performance were obtained for $\alpha=0.81$ and $\varphi_0$ tuned within the range $[0, 3.62]$ rad. This NMSE is higher than that obtained with previous experiments \cite{8,11,13}. However, this is explained by the lack of nodes within the reservoir. For the simulated reservoir with $51$ nodes, NMSE of $0.123\pm0.02$ was obtained. This last result highlights our reservoir capability and shows that performance limitation is only due to current experimental limitations and cannot be attributed to the principle behind it.

\subsection{Nonlinear channel equalization task}

\begin{figure}[htbp]
\centering\includegraphics[width=9cm]{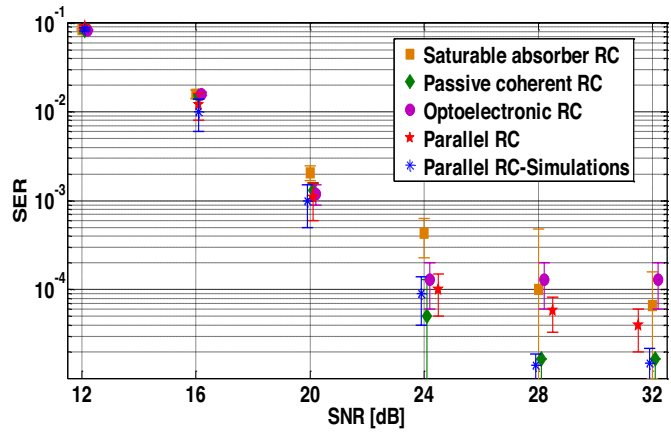}
\caption{ Results obtained for the equalization of the nonlinear channel for signal to noise ratios (SNR) ranging from $12$ to $32$ dB. For each SNR, the symbol error rate (SER) is given with its corresponding error bar over $10$ datasets. The blue and red stars are respectively the simulation ($13$ neurons) and experimental results obtained with the parallel reservoir computer, the magenta circles are the results reported in\cite{8}, the green diamonds are the results presented in \cite{13}, and the orange squares are the results presented in \cite{29}. (Results are obtained for an RF signal of $25$ dBm and $\Omega/2\pi=21$GHz. Parameters $\alpha$ and $\varphi_0$ were scanned respectively in the range of $[0.5, 0.856]$ and $[0,\pi]$ rad).}
\label{Fig3}
\end{figure}

This is a telecommunication applications based task. It was first reported and adapted to the reservoir computing context in\cite{1}. Since then it has been widely used as a benchmark in the reservoir computing community \cite{8,11,13,22}. The aim of this task is, given the channel output $u(n)$, to reconstruct the channel input $d(n)$. The transmitted symbols $d(n)$ are drawn randomly from $\{-3, -1, 1, 3\}$. First, the signal travels through a linear channel resulting in:
\begin{eqnarray}
q(n)&=&0.08d(n+2)-0.12d(n+1)+d(n)+0.18d(n-1)-0.1d(n-2)+0.091d(n-3)\nonumber\\
& &-0.05d(n-4)+0.04d(n-5)+0.03d(n-6)+0.01d(n-7) \ .
\end{eqnarray}
Then, it goes through a noisy nonlinear channel. The resulting noisy outputs u(n) are defined by:
\begin{equation}
u(n)=q(n)+0.036q^2 (n)-0.011q^3 (n)+ {\rm Noise}
\end{equation}
where Noise is an additive zero-mean Gaussian noise which will be adjusted in power to obtain a signal-to-noise ratios ranging from $12$ to $32$ dB. The signal $u(n)$ received at the end of the communication channel is used as the input of the reservoir computer. Then the reservoir will be trained and optimised to generate the original sequence $d(n)$. The reservoir performances are evaluated by calculating the Symbol Error Rate (SER), defining the fraction of the misclassified symbols within the generated $d(n)$ sequence. Simulation and experimental results are reported in Figure \ref{Fig3} and are compared to results obtained in \cite{8,13,29}. We used $3000$ and $6000$ training and testing samples respectively. The SER reported in Figure \ref{Fig3} was evaluated by running the experiment on $10$ different training and test input sequences.

\subsection{Isolated spoken digits recognition}

Finally we use the isolated spoken digits recognition benchmark task to test our reservoir.  The aim of this task is to classify audio sequences. Audio sequences used for this task are taken from the National Institute of Standards and Technology (NIST) TI-46 corpus30. They are composed of five sets of spoken digits between $0$ and $9$ pronounced by five different female speakers. Each of the digits is pronounced $10$ times. These $500$ spoken digits are preprocessed using the Lyon cochlea ear model\cite{31}, and sampled at $12.5$ kHz. Ten classifiers, corresponding to each spoken digit, are trained. If a specific digit is being sent to the reservoir, then the desired output corresponding to this digit is $1$ and $-1$ otherwise. For each of the ten classifiers, outputs are averaged over the sequence length and the actual digit is selected using a winner-take-all approach. The highest averaged classifier is then set to $1$, and all other classifiers are set to $-1$. The performance of the reservoir was evaluated using a cross validation procedure applied over $5$ randomly selected subsets of $100$ words. First, the reservoir was trained over $4$ of the subsets. Then, the last subset is used to test the reservoir. We repeat this procedure $5$ times while rotating the subsets, so that each subset is used once for the test. The results reported here are the average Word Error Rates (WER) and corresponding standard deviations over the $5$ test subsets.

The Lyon ear model produces an $86$ dimensional vectors of outputs $u^l (n)$, $l=1,...,86$. However the present reservoir can only be used with scalar inputs. We therefore transform the output of the Lyon ear model into a scalar input $u(n)$ by taking a linear combination with randomly chosen coefficients, $\mu_l$, among $\{-0.1, 0.1\}$: $u(n)=\sum_{l=1}^{86} \mu_l u^l (n)$.
 
The performance of reservoir is evaluated in terms of Word Error Rate, which report the percentage of the misclassified digits. We obtained a WER of $2.1.\% (\pm 1.2)$ and $2.8\% (\pm 1.5)$ in simulations and experiment respectively. Again, the total number of internal variables is limited to $13$. Our results are significantly worse than those obtained with other architectures adapted to a high number of neurons ($\approx 200$), such as those reported in \cite{7,8,13,23}.  However, our results are comparable to those obtained in \cite{11,32}. We also ran simulations with a modulation amplitude increased artificially until we had $200$ neurons. We then obtained a WER of $0.15\% (\pm 0.1)$, showing again that the performances in our experiment are mainly limited by the number of internal variables. $\alpha$ was tuned between $0.65$ and $0.85$ and $\varphi_0$ was scanned in the region of $[0,\pi]$ rad.

\section{Discussion and Conclusion}

The novel feature of the experimental reservoir computer presented here is the implementation of parallel information processing based on multiplexing neuron states in the frequency domain. This approach allows increasing the total number of neurons without increasing the total processing time. It also allows in principle to increase the number of neurons without increasing the footprint of the reservoir, at least until all the available frequency modes are used. In the present proof of principle demonstration, only a limited number ($13$) of neuron states were used, and the neuron states were read sequentially by tuning the output bandpass filter. In spite of these limitations, good results on several benchmark tasks are obtained and are shown to be comparable to those reported in experiments based on sequential processing of neuron states.
 
The present experiment could be improved in a number of ways. First of all the number of internal states could in principle be increased by increasing the modulation amplitude $m$ of the phase modulator as studied above through simulations. However in practice the maximum value of $m$ is limited. An alternative approach would be to use as a source a frequency comb with spacing $\Omega$, such as a mode locked laser. For instance a mode locked laser with $\Omega/2\pi=10$ GHz and a spectrum spanning $10$ nm would have approximately $100$ neurons. A second improvement would be to decrease the cavity roundtrip time $\tau$. One could consider cavities integrated on a chip for ultrafast applications, in which case $\tau$ in the range $200$ ps to $1$ ns are feasible. Combining a mode locked laser with an on chip reservoir would lead to a $N=100$ neuron reservoir with update time of the input of order $\tau=0,2$ ps to $1$ ns, which would compare favourably with the fastest reservoirs reservoir computers reported sa far \cite{12,14}. Finally one could build an output layer in which all the post-processing is carried out in an analog fashion (see \cite{33} for recent work in this direction). One possibility would be, for instance, to use a frequency demultiplexer combined to an array of photodiodes, one for each frequency sideband, as illustrated schematically in Figure \ref{Fig1B}. Another more attractive possibility would be to carry out output layer (multiplication of the output weights $w_i$ and subsequent summation) all optically, as suggested in a different context in \cite{34,35}.
 
In summary, the present work demonstrates for the first time the real potential of photonic reservoir computers in which all the neuron states are processed simultaneously in parallel at the frequency domain. This approach is technically more elaborate than the time-multiplexed reservoirs that have been most studied up to now. However parallel processing will be essential for all high bandwidth applications. In this sense, our work represents a significant step for the development of future ultrahigh-speed optical signal processors.

\section*{Funding}
The authors acknowledge financial support by Interuniversity Attraction Poles Program (Belgian Science Policy) project Photonics@be IAP P7-35, by the Fonds de la Recherche Scientifique FRS-FNRS, by the Action de Recherche Concert\'ee through grant AUWB-2012-12/17-ULB9.

\end{document}